\documentclass[twocolumn,showpacs,preprintnumbers,amsmath,amssymb]{revtex4}
\usepackage{graphicx}% Include figure files
\usepackage{dcolumn}% Align table columns on decimal point
\usepackage{bm}% bold math

\begin{document}

\preprint{APS/123-QED}

\title{The Electronic Phase Diagram of the Iron-based High $T_c$ Superconductor
Ba(Fe$_{1-x}$Co$_{x}$)$_{2}$As$_{2}$ Under Hydrostatic Pressure ($0\leq x \leq 0.099$)}
% Force line breaks with \\

\author{K. Ahilan$^{1}$, F. L. Ning$^{1}$, T. Imai$^{1,2}$, A. S. Sefat$^{3}$, M. A. McGuire$^{3}$, B. C. Sales$^{3}$, and D. Mandrus$^{3}$}

\affiliation{$^{1}$Department of Physics and Astronomy, McMaster University, Hamilton, Ontario L8S4M1, Canada}
\affiliation{$^{2}$Canadian Institute for Advanced Research, Toronto, Ontario M5G1Z8, Canada}
\affiliation{$^{3}$Materials Science and Technology Division, Oak Ridge National Laboratory, TN 37831, USA}

\date{\today}% It is always \today, today,
             %  but any date may be explicitly specified

\begin{abstract}
%change the abstract below
We report comprehensive resistivity measurements of single crystalline samples of the Ba(Fe$_{1-x}$Co$_{x}$)$_{2}$As$_{2}$ high $T_c$ superconductor under hydrostatic pressure up to 2.75~GPa and over a broad concentration range, $0 \leq x \leq 0.099$.  We show that application of pressure progressively suppresses the SDW transition temperature, $T_{SDW}$, in the underdoped regime ($x \lesssim 0.051$).  There is no sign of pressure-induced superconductivity in the undoped BaFe$_{2}$As$_{2}$ down to 1.8~K, but applied pressure dramatically enhances $T_c$ in the underdoped regime $0.02 \lesssim x\lesssim 0.051$.  The effect of pressure on $T_c$ is very small in the optimally and overdoped regimes $0.082 \lesssim x \lesssim 0.099$. As a consequence, the dome of the superconducting phase extends to $x \lesssim 0.02$ under pressure.  We discuss the implications of our findings in the context of a possible quantum phase transition between the SDW and superconducting phases.
\end{abstract}

% PACS, the Physics and Astronomy
                             % Classification Scheme.
%\keywords{Suggested keywords}%Use showkeys class option if keyword
                              %display desired

\maketitle

%\section{\label{sec:level1}First-level heading:\protect\\ The line
%break was forced \lowercase{via} \textbackslash\textbackslash}

\section{\label{sec:level1}Introduction}
The superconducting mechanism of the new iron-based high $T_c$ superconductors \cite{Kamihara} is highly controversial. Among the key questions which remain unsolved is whether the conventional phonon mechanism is responsible for the superconductivity. Given that the ground state of the undoped parent phases, such as RFe$_{2}$As$_{2}$ (R = Ba, Sr, Ca), is magnetically ordered in a commensurate SDW (Spin Density Wave) state \cite{Huang, Kitagawa, Jesche, Ronning, Ni_CaFeAs}, it is conceivable that spin fluctuations may be playing a role as a glue of Cooper pairs. Unlike the high $T_c$ cuprate superconductors, however, the magnetically ordered ground state of the undoped parent phases is not a Mott insulating state, and the electrical resistivity $\rho$ remains finite in the SDW state \cite{Kamihara, Rotter1}. As little as 2 to 4 percent of electron doping into the FeAs layers alters the nature of the SDW order, as evidenced by the dramatic changes of the $^{75}$As and $^{59}$Co NMR lineshapes in the ordered state of Ba(Fe$_{1-x}$Co$_{x}$)$_{2}$As$_{2}$ \cite{Ning3}.  The exact nature of the SDW phase in the presence of doped electrons is not understood very well, but the large distributions of the static hyperfine magnetic field observed by NMR are not consistent with a homogeneous, commensurate SDW state \cite{Ning3}.  Upon further increasing the level of electron doping to the optimal doping level of 6 to 8 \%, a high $T_c$ phase emerges with $T_{c} \lesssim 23$~K \cite{Sefat1, Ning2, Ni, Chu1, Wang1}.  

Besides doping, it also turns out that applied pressure can induce superconductivity with $T_c$ as high as $\sim 29$~K  in the undoped parent phases of RFe$_{2}$As$_{2}$ \cite{Alireza, Fukazawa, Torikachvili1, Park, Igawa, Kotegawa}.  The existence of the pressure induced superconducting phase indicates that subtle changes and/or contractions of the structure can switch on superconductivity from a SDW phase.  However, the mechanism of pressure induced superconductivity is very poorly understood, and more detailed studies are required to clarify the effects of applied pressure on the electronic properties of iron-based high $T_c$ superconducting systems.  Most of the past experimental studies of these pressure effects, however, have focused on the optimally doped superconducting phase, or on pressure induced superconductivity in the undoped parent phase (see \cite{Chu2} for a review).  Only limited experimental studies have been reported for the interplay between the amount of doping {\it and} pressure on the RFe$_{2}$As$_{2}$ systems \cite{Ahilan}.  In this paper, we will present comprehensive resistivity measurements under hydrostatic pressure for Ba(Fe$_{1-x}$Co$_{x}$)$_{2}$As$_{2}$ single crystals over a wide range of Co concentrations from the undoped ($x = 0$) to overdoped regimes up to $x=0.099$.  Unlike earlier reports of the observation of superconductivity under pressures applied by anvil cells \cite{Alireza,Fukazawa}, we do {\it not} observe superconductivity in the undoped BaFe$_{2}$As$_{2}$ at least up to 2.75~GPa.  On the other hand, we do find that applied pressure strongly enhances $T_c$ in the underdoped regime $0.02 \lesssim x \lesssim 0.051$.  The pressure effect on $T_c$ is very weak in the optimum and overdoped regimes, hence the {\it dome} of the superconducting region in the phase diagram extends toward $x=0$ under hydrostatic pressures. 

The rest of this paper is organized as follows:  in Section II, we describe experimental details.  Our experimental results in ambient pressure and under hydrostatic pressure are described in Sections III and IV, respectively, followed by summaries and conclusions in Section V.

%============
\section{\label{sec:level1}Experimental Methods}
We grew Co-doped BaFe$_{2}$As$_{2}$ single crystals based on FeAs self-flux methods \cite{Sefat1}.  The samples were cleaved and cut into small pieces with typical dimensions of 2 mm  $\times$ 1 mm $\times$ 0.15 mm for electrical transport measurements.   We applied high pressures of up to 2.75 GPa using a compact hybrid pressure cell with a BeCu outer jacket and a NiCrAl inner core.  Daphene oil 7373 and 99.99$\%$ purity Sn were used as a pressure transmitting medium and a pressure calibrating gauge, respectively.  Sample contacts were made using silver epoxy for conventional four-lead ac-resistivity measurements.  We employed a highly flexible, home-made ac-resistivity measurement rig to achieve high accuracy in the resistivity measurements.  The high pressure cell was placed in a vacuum canister with helium exchange gas.  All of the measurements reported in this paper were carried out while warming up the sample from the base temperature of 1.8~K.  In order to ensure that thermal equilibrium was reached properly, we stabilized the temperature of the system before conducting the resistivity measurement at each temperature, instead of continuously ramping the temperature.  We confirmed that measurements carried out in the warming cycle agree well with those in the cooling cycle for all of the measurements.  We exercised these precautions because the heat capacity of the high pressure cell is rather large, and the Cernox temperature sensor is attached to the exterior of the high pressure cell.   We estimate the upper bound for the potential inaccuracy in sample temperature at 0.5~K.

In this paper, we present the details of the resistivity measurements primarily for $x=0$, 0.02, 0.051, and 0.097.  We refer readers to Ref. \cite{Ahilan} for the additional details of measurements in $x=0.04$ and 0.082.  

%============
\section{\label{sec:level1} Results in Ambient Pressure}
We begin our discussions with a summary of the resistivity data $\rho_{ab}$ in ambient pressure, $P=0$, shown in Fig.1a. In the undoped sample with $x=0$, $\rho_{ab}$ decreases suddenly below $T_{SDW} = 135$~K, in agreement with earlier reports \cite{Rotter2,Huang}. The cause of this dramatic change has been identified as a first order SDW phase transition accompanied by a structural phase transition from a high temperature tetragonal to low temperature orthorhombic structure \cite{Huang}.  We note that $\rho_{ab}$ would increase below $T_{SDW}$ if SDW energy gaps open for all branches of bands crossing the Fermi energy \cite{Jerome}.  Instead, $\rho_{ab}$ actually decreases below $T_{SDW}$ in the undoped sample.

\begin{figure}
\centering
\includegraphics[width=3in]{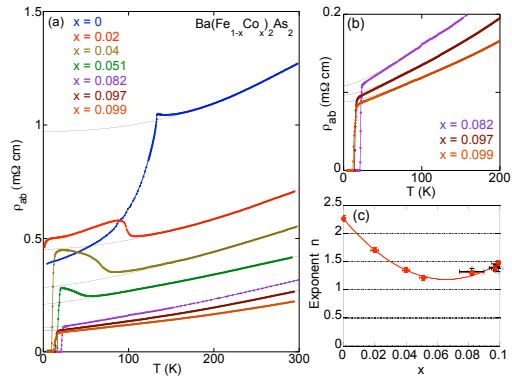}% Here is how to import EPS art
\caption{\label{Fig1:epsart}(a) The temperature dependence of the in-plane resistivity $\rho_{ab}$ of Ba(Fe$_{1-x}$Co$_{x}$)$_{2}$As$_{2}$ single crystals in ambient pressure, $P=0$.  Dotted curves are the best fits to an empirical relation, $\rho_{ab} = A + BT^{n}$, above 170~K, where $A$ and $B$ are constants.  The concentration dependence of $n$ is summarized in panel (c).  (b) The same data shown on a magnified scale for the optimal ($x = 0.082$) and overdoped ($x = 0.097, 0.099$) regimes.  Notice that the extrapolation of the aforementioned fits markedly deviate from the data near $T_c$, where $\rho_{ab}$ shows T-linear behavior.  (c) Filled circles : the exponent $n$ obtained from the fit of $\rho_{ab}$ above 170~K.  Open diamonds : the exponent $n$ obtained from a more global fit above $T_c$ in the superconducting samples which do not show an SDW transition ($x = 0.082$, 0.097, 0.099).}
\end{figure}

\begin{figure}
\centering
\includegraphics[width=3in]{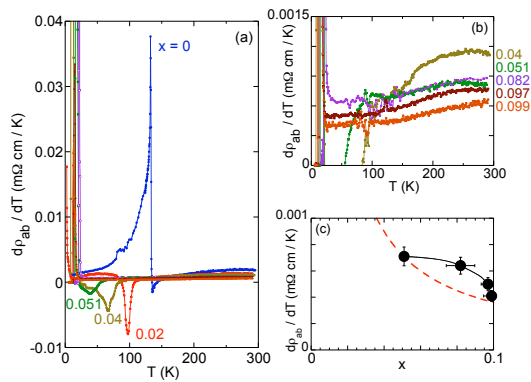}% Here is how to import EPS art
\caption{\label{Fig3:epsart}(a) The temperature dependence of the derivative of the in-plane resistivity, $d\rho_{ab}/dT$, for Ba(Fe$_{1-x}$Co$_{x}$)$_{2}$As$_{2}$ single crystals with \emph{x} = 0, 0.02, 0.04, and 0.051 in ambient pressure, $P=0$.  (b) $d\rho_{ab}/dT$ for x=0.04 and above in a magnified scale.  Notice that upon cooling, $d\rho_{ab}/dT$ levels off toward  a constant value near $\sim 100$~K for $x=0.051$ and above, implying that $\rho_{ab} \propto T$. (c) Filled circles : the constant slope $d\rho_{ab}/dT$ in the $\rho_{ab} \propto T$ regime from $T_c$ to $\sim 100$~K as obtained from the results in panel (b) for the superconducting samples with $x=0.051$ and above. Dashed curve represents $d\rho_{ab}/dT \sim 1/x$, and does {\it not} account for the $x$-dependence of $d\rho_{ab}/dT$ near $T_c$.
} 
\end{figure}

Once we dope a few percent of Co into the Fe sites, however, the abrupt drop of $\rho_{ab}$ is no longer observable, and $\rho_{ab}$ exhibits a step-like increase \cite{sefat2, Ahilan}. We proposed earlier that the temperature derivative $d\rho_{ab}/dT$ of the resistivity data permits us to characterize the step-like anomaly \cite{Ahilan}.  We show the summary of  the temperature dependence of $d\rho_{ab}/dT$ in Fig.2. The minimum of $d\rho_{ab}/dT$ is clearly observable at $T_{SDW}=100\pm1$~K for $x=0.02$.  The justification for identifying the minimum of $d\rho_{ab}/dT$ as $T_{SDW}$ is that our $^{75}$As and $^{59}$Co NMR measurements for the same batch of crystals reveal typical signatures of a second order magnetic phase transition at the same temperature, including divergent behavior of the nuclear spin-lattice relaxation rate $1/T_{1}$ and the onset of broadening of the NMR lineshapes \cite{Ning2,Ning3}.  The enhancement of $1/T_{1}$ originates from the critical slowing down of the low frequency components of spin fluctuations  toward a magnetic phase transition, while the NMR line broadening is due to the growth of spontaneous magnetization below a magnetic phase transition.

For higher doping levels with $x=0.04$ and 0.051, the resistivity upturn is less pronounced.  By applying the same criterion based on the minimum of $d\rho_{ab}/dT$, we determined the SDW transition temperature as $T_{SDW} = 66\pm1$~K for $x=0.04$ and $T_{SDW} = 40\pm1$~K for $x=0.051$, respectively.

We also observe a clear signature of a resistive superconducting transition for samples with $x=0.04$ or above.  For example, the $x=0.04$ sample exhibits the onset of superconductivity at $T_{c}=11.0\pm0.5$~K.  In what follows, we define the superconducting transition temperature $T_c$ as the temperature where $\rho_{ab}$ decreases by 10~\% from the extrapolated linear behavior of $\rho_{ab}$ from higher temperature. Interestingly, even the $x=0.02$ crystal shows a slight decrease of $\rho_{ab}$ below 4~K.  The observed decrease is very subtle in ambient pressure. However, application of hydrostatic pressures dramatically enhances the onset temperature of the drop in $\rho_{ab}$ to as high as 10.5~K, and $\rho_{ab}$ reaches almost zero at 1.8~K in 2.4~GPa, as discussed in Section IV.  We also confirmed that application of a 9~T magnetic field suppresses the onset of the resistivity drop, while NMR measurements \cite{Ning3} reveal no additional magnetic anomaly around $\sim 10$~K or below. We therefore conclude that even the lightly electron-doped $x=0.02$ crystal has a resistive superconducting transition.

We summarize the concentration $x$ dependence of $T_{SDW}$ and $T_c$ in the electronic phase diagram shown in Fig. 3.  The Fe$_{1-x}$Co$_{x}$As layers in the underdoped regime, $x \lesssim0.051$, undergo successive SDW and superconducting phase transitions.  The $T_c$ reaches a maximum value of $\sim 23.6$~K in the optimally doped region for $0.06 \lesssim x \lesssim 0.082$.  Although there is no SDW transition in the optimally doped regime, the NMR spin-lattice relaxation rate $1/T_{1}T$ is enhanced above $T_c$, providing evidence for the presence of residual antiferromagnetic spin fluctuations \cite{Ning1,Ning2}. Once we enter the overdoped regime, $T_c$ begins to decrease, and the NMR data no longer show evidence for enhanced antiferromagnetic spin fluctuations near $T_c$ \cite{Ning2}. We will discuss the implications of the phase diagram in Section IV combined with the results of our measurements under hydrostatic pressure.

\begin{figure}
\centering
\includegraphics[width=3in]{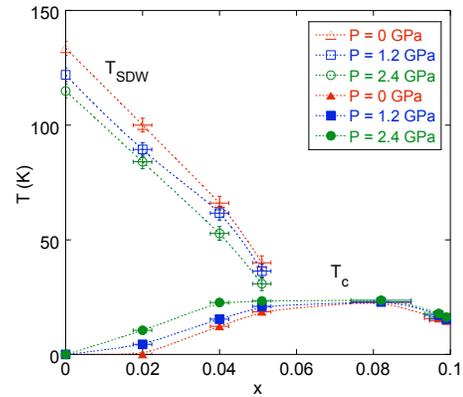}% Here is how to import EPS art
\caption{\label{Fig2:epsart} The \emph{x-T} electronic phase diagram of magnetism and superconductivity for Ba(Fe$_{1-x}$Co$_{x}$)$_{2}$As$_{2}$ under pressures of $P= 0$, 1.2, and 2.4~GPa. Open and filled symbols represent $T_{SDW}$ and $T_{c}$, respectively.}
\end{figure}

\begin{figure}
\centering
\includegraphics[width=3in]{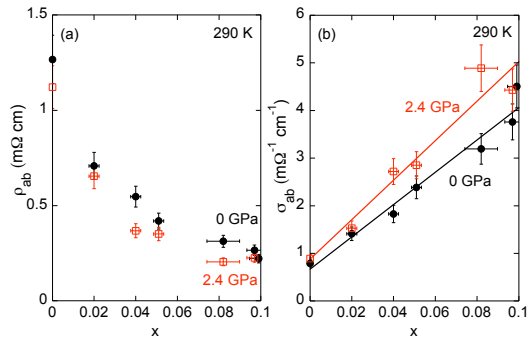}% Here is how to import EPS art
\caption{\label{Fig4:epsart} (a) The concentration dependence of the in-plane resistivity $\rho_{ab}$ at 290~K for $P=0$ (filled circles) and $P=2.4$~GPa (open squares).  (b) The conductivity $\sigma_{ab} = 1/\rho_{ab}$ at 290~K.  Solid lines represent the best linear fits to the data.
}
\end{figure}

Among many puzzling aspects of the electronic properties of electron-doped Ba(Fe$_{1-x}$Co$_{x}$)$_{2}$As$_{2}$ is the temperature $T$ and concentration $x$ dependencies of  $\rho_{ab}$.  One can fit the overall temperature dependence of $\rho_{ab}$ above $T_{SDW}$ and $T_c$ to a power-law behavior, $\rho_{ab} = A + BT^{n}$, with a constant background $A$ \cite{Ahilan}.  The exponent $n$, as determined from the fit in the temperature range above 170~K so that we could apply the same fitting criterion for all samples, shows only a mild concentration dependence, $n = 1.3 \sim 1.5$, above $x =0.04$, as summarized in Fig. 1c.  We note that the value of $n$ does not depend very strongly on the fitting range.  For example, even if we extend the fitting range down to $T_c$ in the superconducting samples with $x \geq 0.082$, the deduced values (shown by open diamonds in Fig.1c) are about the same.  The observed exponent is very close to 4/3 or 3/2, typical values observed in some heavy Fermion systems \cite{Sarrao} or overdoped high $T_c$ cuprates \cite{Takagi}.  It is not clear, however, if the assumption of the presence of a large temperature independent background resistivity $A$ is justifiable, especially in the underdoped regime. We also note  that the extrapolation of the fit below 170~K poorly reproduces the data below $\sim 100$~K in all of the superconducting samples, because $\rho_{ab}$ asymptotically approaches T-linear behaviors, as evidenced by constant slopes below $\sim 100$~K in Fig.2b.  This T-linear behavior is not consistent with canonical Fermi liquid behavior, $\rho_{ab} \sim T^{2}$.  We recall that $^{75}$As NMR results in the T-linear regime do not satisfy the Korringa law, $1/T_{1}T \propto [K_{spin}]^{2}$, expected for a Fermi liquid either \cite{Ning1,Ning2} ($K_{spin}$ is the spin contribution to the NMR Knight shift, which is proportional to the uniform spin susceptibility). 

In order to quantify the systematic variation of $\rho_{ab}$, we summarize the $x$ dependence of $\rho_{ab}$ at 290~K in Fig. 4a.  $\rho_{ab}$ decreases monotonically with $x$.  To better understand the systematic trend, we also plot the in-plane conductivity $\sigma_{ab}$~($= 1/\rho_{ab}$) in Fig. 4b.  The latter suggests that $\sigma_{ab}$ at 290~K increases in proportion to $x$, $\sigma_{ab}(x) = \sigma_{ab}(0) + Cx$ where $C$ is a constant.  We found that the concentration dependence of $\sigma_{ab}$ at a fixed temperature above 150~K shows analogous linear dependence on $x$ as long as all samples remain paramagnetic.  It is well known that similar linear $x$ dependence of $\sigma_{ab}$ was also observed in the high $T_c$ cuprates for a broad concentration range \cite{Takagi,Ando1,Ando2}.

In the case of the high $T_c$ cuprate La$_{2-x}$Sr$_x$CuO$_4$, $\rho_{ab}$  exhibits T-linear behavior over a broad temperature range up to as high as $\sim 1000$~K and for a broad hole concentration range from $x=0.01$ to $0.22$ \cite{Ando2}.  This T-linear behavior may be caused by, among other possibilities, the quantum criticality \cite{Chubukov}.  The persistence of the T-linear behavior to such high temperatures implies that the fundamental energy scale which dictates the electronic properties of high $T_c$ cuprates is large (e.g. the Cu-Cu superexchange interaction $J$ is as large as $\sim 1500$~K in undoped La$_2$CuO$_4$).  In contrast, in the present case, the slope $d\rho_{ab}/dT$ is constant (hence  $\rho_{ab} \sim T$) only below $\sim 100$~K and only for $x \gtrsim 0.051$, as shown in Fig.2b. The fact that the T-linear behavior breaks down above $\sim 100$~K suggests that the fundamental energy scale of the electronic properties of FeAs layers is relatively low.

Another important distinction is the concentration dependence of  the slope.  In cuprates, the slope is roughly inversely proportional to the doped hole concentration, i.e. $d\rho_{ab}/dT \sim 1/x$ in the T-linear regime \cite{Ando2}. This implies that each hole in the CuO$_{2}$ planes of the high $T_c$ cuprates contributes to the in-plane conductivity $\sigma_{ab}$~($= 1/\rho_{ab}$) by the same amount, regardless of the level of doping.  In the present case, however, the low temperature slope in the T-linear region does not vary as $\sim 1/x$, as shown in the Fig.2c.  Instead, $d\rho_{ab}/dT$ is roughly constant in the optimally doped regime, and begins to decrease very rapidly once we enter the overdoped regime above $x=0.082$.  In fact, upon further increasing $x$, BaCo$_{2}$As$_{2}$ with $x=1$ has a Fermi-liquid-like ground state, and satisfies $\rho_{ab} \sim T^{2}$ below $\sim 70$~K \cite{Sefat3, Sefat4}.  The latter implies that $d\rho_{ab}/dT$ asymptotes to zero with decreasing temperature for $x=1$.  Recent studies suggest that this crossover into the Fermi-liquid-like ground state with  $\rho_{ab} \sim T^{2}$ takes place around $x=0.2$ \cite{Ni,Chu1,Wang1}.

%============
\section{\label{sec:level1} Pressure Effects}

In Fig.5, we summarize representative results of resistivity measurements under hydrostatic pressures for undoped $x=0$, lightly doped $x=0.02$, underdoped $x=0.051$, and overdoped $x=0.097$ samples.   We refer readers to our earlier report for the details of the measurements on underdoped $x=0.04$ and optimally doped $x=0.082$ samples \cite{Ahilan}.  In all cases, $\rho_{ab}$ does not show major qualitative changes under hydrostatic pressure.  The magnitude of $\rho_{ab}$ decreases by $\sim 20$\% from 0~GPa to $\sim 2.4$~GPa, but the empirical relation for the concentration dependence, $\sigma_{ab}(x) = \sigma_{ab}(0) + Cx$, still holds with a $\sim 20$\% larger value of $C$, as shown in Fig.4b.  The exponent $n$ from the  fit to $\rho_{ab} = A + BT^{n}$ is also comparable between $P=0$ and 2.4~GPa \cite{Ahilan}.

\begin{figure}
\centering
\includegraphics[width=3.5in]{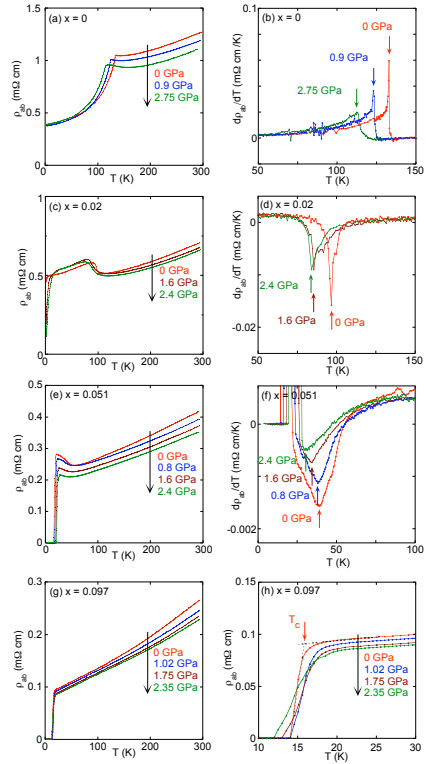}% Here is how to import EPS art
\caption{\label{Fig5:epsart} Left panels : In-plane resistivity $\rho_{ab}$ under pressure for various values of Co concentration $x$. The upper right panels (b), (d), and (f) show the derivative of the in-plane resistivity $d\rho_{ab}/dT$.  The bottom right panel (h) shows the resistive superconducting transition for $x=0.097$ under various pressures.}
\end{figure}

How does hydrostatic pressure affect the phase transition temperatures $T_{SDW}$ and $T_c$?  In the case of undoped BaFe$_{2}$As$_{2}$, our $\rho_{ab}$ data in Fig.5a and its derivative $d\rho_{ab}/dT$ in Fig.5b clearly show progressive suppression of $T_{SDW}$ from 135~K in 0~GPa to 114~K in 2.75~GPa. The extremely sharp peak of $d\rho_{ab}/dT$ at $T_{SDW}=135$~K becomes broader under pressure, and the sharp peak is no longer observable in 2.75~GPa.  This may be an indication that the first order nature of the phase transition at $T_{SDW}$ in 0~GPa becomes gradually weaker under pressure, and quite possibly the SDW transition may become second order.  However, we cannot entirely rule out an alternate scenario in which the applied pressure has a mild distribution due to the freezing of the pressure medium etc., and therefore $T_{SDW}$ itself has a small distribution under pressures, especially at $P=2.75$~GPa.  A distribution of $T_{SDW}$ over $\sim 5$~K would easily mask the sudden nature of the first order transition, and make the transition appear to be of the second order.

Another important aspect of our $\rho_{ab}$ data for  the undoped BaFe$_{2}$As$_{2}$ is that we find no hint of a resistive superconducting transition up to at least 2.75~GPa.   In contrast with our results, earlier SQUID measurements detected diamagnetic Meissner signals of a superconducting transition with $T_c$ as high as 29~K in BaFe$_{2}$As$_{2}$ above a critical pressure $P_{c}\sim 2.8$~GPa applied by diamond anvil cell \cite{Alireza}.  A subsequent report on resistivity measurements in non-hydrostatic pressure applied by a Bridgman cell also detected strong suppression of resistivity above a comparable $P_c$, although  zero resistivity was never observed \cite{Fukazawa}.  In the present case, we cannot rule out the possibility that  our maximum hydrostatic pressure of 2.75~GPa ($< P_c$) is somewhat too low to induce superconductivity.    Our compact hydrostatic high pressure cell risks damage, or even a catastrophic failure, at $\sim 3$~GPa or higher.  Accordingly, we have not explored the pressure range above 2.8~GPa.   However, it is worth pointing out that our $\rho_{ab}$ data in 2.75~GPa shows a robust signature of an SDW transition at $T_{SDW}=114$~K; it seems highly unlikely that bulk superconductivity suddenly sets in under hydrostatic pressure at  $P_{c}\sim 2.8$~GPa unless a structural phase transition takes place between 2.75 and 2.8~GPa.  

\begin{figure}
\centering
\includegraphics[width=3.5in]{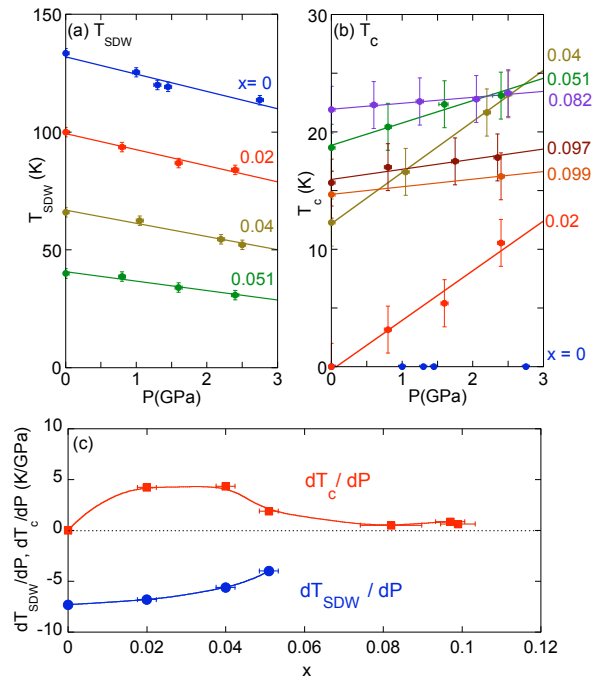}% Here is how to import EPS art
\caption{\label{Fig4:epsart} The pressure $P$ dependence of (a) $T_{SDW}$, and (b) $T_c$.  Solid lines are the best linear fits.  (c) The pressure coefficient $dT_{c}/dP$ and $d T_{SDW}/dP$ as a function of the Co concentration $x$.  Solid curves are guides for the eyes.}
\end{figure}

Next, we turn our attention to the interplay between Co doping and applied pressure. We summarize $T_{SDW}$ and $T_c$ as a function of hydrostatic pressure in Fig. 6a, and b, respectively.  Fig.6c summarizes the pressure coefficient, $dT_{SDW}/dP$ and $dT_{c}/dP$, for various Co doping levels based on the linear fits of the data points in Fig.6a and b.  The results in Fig.6a and c clearly establish that the pressure induced suppression of $T_{SDW}$ becomes progressively weaker as we increase the Co concentration from $x=0$ to 0.051. On the other hand, Fig.6b and c show that hydrostatic pressure always enhances $T_c$, but the sensitivity of $T_c$ on pressure depends strongly on the Co concentration $x$.  $dT_{c}/dP$ reaches  as large as +4.3~K/GPa for $x=0.02 \sim 0.04$, but $dT_{c}/dP$ decreases to $\leq +1$~K/GPa in the optimum and overdoped regimes.   It is not clear why the pressure effect on $T_c$ becomes so weak for $x=0.082$ or above.  Our observation of $dT_{c}/dP > 0$ in both $x = 0.04$ and 0.082  also defies the conventional wisdom based on the Ehrenfest relation of $T_c$ as estimated by thermal expansion data \cite{Budko}.  Another remarkable point from Fig. 6c is that the magnitude of {\it both} $dT_{SDW}/dP$ and $dT_{c}/dP$ decrease strongly near $x \sim 0.06$. In other words, the effects of pressure on $T_c$ and $T_{SDW}$ become weak near $x \sim 0.06$. These trends suggest the existence of a crossover in the electronic properties near $x \sim 0.06$.  This finding may be related to a recent report on the crystal structure which showed that Ba(Fe$_{1-x}$Co$_{x}$)$_{2}$As$_{2}$ does not undergo a high temperature tetragonal to low temperature orthorhombic structural phase transition in the concentration range $x \gtrsim 0.06$ \cite{Lester}.

In passing, it is worth noting that the pressure effects on $T_c$ do not seem to obey a simple universal behavior in other systems either, and what dictates the pressure induced change of $T_c$ is not clear.   For example, in the case of the (Ba$_{1-x}$K$_{x}$)Fe$_{2}$As$_{2}$ system with x = 0.45, $T_c$ decreases smoothly with pressure up to 2~GPa at a rate $dT_{c}/dP\sim -2.1$~K/GPa \cite{Torikachvili2}. On the other hand, for the La(O$_{1-x}$F$_{x}$)FeAs system $T_{c}\sim28$~K sharply increases to 43~K with pressure up to 3~GPa at an average rate $dT_{c}/dP \sim +5$~K/GPa, but application of higher pressure suppresses $T_c$ \cite{Takahashi}.

%=================
\section{\label{sec:level1}Summary and Conclusions}
By linearly interpolating data points in Fig.~6a and b at $P = 1.2$ and 2.4~GPa, we construct the electronic phase diagram of Ba(Fe$_{1-x}$Co$_{x}$)$_{2}$As$_{2}$ under pressure  at $P = 1.2$ and 2.4~GPa in Fig.~3.  Application of a hydrostatic pressure of 2.4~GPa suppresses $T_{SDW}$ by $10 \sim 19$~K for all samples with a SDW transition.  On the other hand, applied pressures enhance$T_c$ dramatically only in the underdoped region, while affecting $T_c$ little in the optimum and overdoped regimes.  Accordingly, the optimally doped regime with $T_{c}(2.4GPa) \sim 23.6$~K extends to as low as $x = 0.04 \sim 0.051$.  Also notice that the optimally doped region emerges when the magnetic phase boundary $T_{SDW}$ intersects the dome of the superconducting phase near $x \sim 0.05$ in $P=2.4$~GPa.  In the case of ambient pressure, the intersection is located at a somewhat higher value near $x \sim 0.06$ \cite{Ning2,Ni,Chu1,Wang1}.

One can take several different views on the phase diagram in Fig.3.  One possible scenario is that the SDW and superconducting phases compete each other.  In this viewpoint, one can attribute the
extension of the optimum $T_c$ region to $x = 0.04 \sim 0.05$ in 2.4~GPa as a consequence of the suppression  of the SDW instability by pressure.  One can also take a completely opposite viewpoint; notice that if we traverse the phase diagram near $T=0$ from the superconducting phase $x \sim 0.1$ toward $x=0$, low frequency antiferromagnetic spin fluctuations as a function of $x$ would diverge at $x_{c}\sim 0.06$ in $P=0$~GPa and  $x_{c}\sim 0.05$ in $P=2.4$~GPa when we hit the boundary with the SDW  phase, i.e. {\it a quantum phase transition} at $x_{c}$ from the superconducting to SDW ground state.  In fact, our earlier NMR measurements showed enhancement of antiferromagnetic spin fluctuations  for lower doping level $x$ \cite{Ning2}, and confirmed such trends.

In this second scenario, our phase diagram in Fig. 3 might imply that enhanced quantum spin fluctuations near $x_{c}$ are the key to the superconducting mechanism.  We recall that an analogous scenario involving quantum criticality has been debated extensively  in the context of high $T_c$ cuprates since the early 1990's \cite{Imai1, Chubukov, Sokol}, and more recently in the context of pressure induced superconductivity in heavy Fermions \cite{Saxena, Mathur}. The recent finding that application of hydrostatic pressure enhances both $T_c$ \cite{Mizuguchi, Medvedev, Margadonna} and antiferromagnetic spin fluctuations in FeSe \cite{Imai2} renders additional support to this second scenario, because spin fluctuations would be suppressed by pressure if superconductivity genuinely competes with the SDW instability.

On the other hand, one may need to be somewhat cautious in the debate over the cooperation or competition between superconductivity and SDW in the present case of Ba(Fe$_{1-x}$Co$_{x}$)$_{2}$As$_{2}$, because the structural phase boundary between the tetragonal and orthorhombic phases terminates near $x=0.06$ \cite{Lester}.  We cannot rule out the possibility that subtle changes in the crystal structure turn off the SDW order rather suddenly and switch on superconductivity.  In this third scenario, $x_c$ decreases from $\sim 0.06$ in 0~GPa to  $\sim 0.05$ in 2.4~GPa as a consequence of the shift of the tetragonal-orthorhombic structural boundary to $x \sim 0.05$ under pressure.  Further structural studies under pressure are required to test the scenario.

%============
\section{\label{sec:level1}Acknowledgment}
The work at McMaster was supported by NSERC, CFI, and CIFAR.  Research at ORNL was sponsored by Division of Materials Sciences and Engineering, Office of Basic Energy Sciences, U.S. Department of Energy.\\

%\bibliography{Ahilan_PRB}% Produces the bibliography via BibTeX.

\end{document}